\documentclass[a4paper,11pt]{article}

\usepackage[backend=bibtex,style=numeric-comp,natbib=true,sorting=none,maxbibnames=10 ]{biblatex} 
\usepackage[autostyle=true]{csquotes}

\usepackage{amsmath, amsthm, amssymb, amsfonts}
\usepackage{graphicx} 
\usepackage{hyperref}
\usepackage{cases}
\usepackage{braket}
\usepackage{caption}
\usepackage{float}
\usepackage{empheq}

\usepackage{color}
\definecolor{Blue}{rgb}{0.00, 0.00, 1.00}
\definecolor{Red}{rgb}{1.00, 0.00, 0.00}
\definecolor{Green}{rgb}{0.00, 0.70, 0.00}
\newcommand{\red}{\color{Red}}
\newcommand{\blue}{\color{Blue}}

\hypersetup{
    colorlinks=true,       
    linkcolor=red,          
    citecolor=blue,        
    filecolor=magenta,      
    urlcolor=cyan           
}

\numberwithin{equation}{section}

\newcommand{\nn}{\nonumber}
\newcommand{\be}{\begin{equation}}
\newcommand{\ee}{\end{equation}}
\newcommand{\Ai}{{\rm Ai}}
\newcommand{\al}[1]{\begin{align}#1\end{align}}
\newcommand{\simh}{\underset{\hbar \to 0}{\simeq}}

\usepackage{titlesec}
\titleformat{\section}{\large\bf}{\thesection}{1em}{}
\titleformat{\subsection}[runin]{\bf}{\thesubsection}{1em}{}

\usepackage{authblk}

\makeatletter
\newcommand\appendix@section[1]{%
  \refstepcounter{section}%
  \orig@section*{Appendix \@Alph\c@section: #1}%
  \addcontentsline{toc}{section}{Appendix \@Alph\c@section: #1}%
}
\let\orig@section\section
\g@addto@macro\appendix{\let\section\appendix@section}
\makeatother

\title{ \bf  Quantum propagating front and the edge of the Wigner function}

\author{Gabriel Gouraud}
\affil{\normalsize Laboratoire de Physique de l'\'Ecole Normale Sup\'erieure, ENS, Universit\'e PSL, CNRS, Sorbonne Universit\'e, Universit\'e de Paris, 75005 Paris, France}

\begin{document}

\maketitle
\begin{abstract}
In the first part of the article, we study one-dimensional noninteracting fermions in the continuum and in the presence of the repulsive inverse power law potential, with an emphasis on the Wigner function in the semiclassical limit. In this limit, the Wigner function exhibits an edge called the Fermi surf that depends only on the classical one-particle Hamiltonian. Around the Fermi surf, under a well-defined semiclassical limit, the Wigner function can be expressed in terms of Airy functions which yield a smooth matching between the two regions delimited by the Fermi surf. 
In the second part of the article, the system is prepared in the ground state of the inverse power law potential where only the left half line is filled with fermions. Then the potential is switched off, resulting in the emergence of a propagating quantum front. We show that the power law decay of the pre-quench potential that separates the left and right half systems leads to the emergence of the Airy kernel (well known in Random Matrix Theory) at the quantum front in the long-time limit. This also comes with anomalous diffusive spreading around the front. 
\end{abstract}

\section{Introduction}
Historically, the study of quantum many-body systems has been pursued under the equilibrium or maximal entropy hypothesis. In this context, the Wigner function \cite{wigner1997quantum} has proven particularly useful as a phase space approach to quantum mechanics. The Wigner function was considered by Berry \cite{berry1977semi} in the case of the semiclassical limit of a single-particle state. It has been employed in various contexts \cite{case2008wigner}, including quantum optics \cite{walls2008quantum}, the modeling of optical devices \cite{bazarov2012synchrotron}, and quantum information \cite{douce2013direct}. One natural question is whether the Wigner function can be measured. This has been accomplished with quantum state tomography \cite{smithey1993measurement} and trapped atom setups \cite{leibfried1996experimental}.\par
The Wigner function of one-dimensional trapped fermions in their ground state was first studied by Balazs and Zipfel \cite{balazs1973quantum}. In the semiclassical limit, the Wigner function exhibits an edge called the Fermi surf. For one dimensional systems, the behavior of the Wigner function around the Fermi surf was analysed both in and out of equilibrium \cite{bettelheim2011universal,bettelheim2012quantum}. This study was latter extended to dimensions lager than one and non-zero temperature set up \cite{dean2018wigner,dean2019nonequilibrium}. At equilibrium trapped Fermi gases has raised interest both on the experimental and theoretical side \cite{bloch2008many,giorgini2008theory}. Despite their simplicity, noninteracting Fermi gases have been shown to exhibit rich universal behaviors \cite{kohn1998edge,eisler2013universality,dean2015finite,dean2015universal,dean2017statistics,dean2018wigner,calabrese2011entanglement,marino2014phase,dean2019noninteracting,de2021wigner}. The latter also have multiple connections with the eigenvalues of random matrices, see \cite{dean2016noninteracting} for a recent review. In particular, there is an exact connection between the positions of noninteracting fermions trapped in a harmonic potential and the eigenvalues of the Gaussian Unitary Ensemble (GUE) \cite{eisler2013universality,calabrese2015random,dean2019noninteracting,marino2014phase}.\par

On the other side, non-equilibrium quantum mechanics has been a rapidly developing topic, particularly with the aim of addressing transport in mesoscopic materials. This was done by the so called Landauer-Büttiker formalism \cite{landauer1989johnson,buttiker1992scattering,blanter2000shot}. This understanding is well-established for the transport of noninteracting electrons between different channels separated by a scattering defect. In this case, the study was extended to Full Counting Statistics (FCS) \cite{levitov1993charge}, which allows for a complete characterization of current noise.

More recently, the theoretical and experimental attention has focused on investigating the full dynamics using quench setups \cite{polkovnikov2011colloquium,calabrese2006time,platini2007relaxation,calabrese2007quantum,calabrese2011quantum}. There are many kinds of quenches, for example, the interaction can be changed \cite{mitra2011mode}, it can be studied in different contexts whether the system is a Luttinger liquid \cite{cazalilla2006effect,foini2015nonequilibrium,iucci2009quantum,foini2017relaxation}, is integrable \cite{kinoshita2006quantum,alba2017entanglement,essler2016quench}, or when Conformal Field Theory can be used \cite{collura2015quantum,foini2015nonequilibrium,ruggiero2021quenches,bernard2012energy,bernard2015non,capizzi2023entanglement,stephan2011local}. Specifically, substantial focus has been on one-dimensional quenches called bipartition protocol, where two halves of the system are prepared with different densities. The system then evolves freely, inducing a long-time stationary current from the higher to the lower density region \cite{antal1999transport, eisler2013full, Collura_2014}. A scattering defect can be added to this quench \cite{schonhammer2007full,gamayun2020fredholm,gamayun2023landauer,gouraud2022quench,gouraud2022stationary}, such that the large time limit yields the stationary statistics found with the Landauer Büttiker formalism \cite{landauer1989johnson,levitov1993charge}. However, in this discussion, we focus on the case without defects.

In such a step-like initial density quench, an interesting phenomenon arises: the emergence of a propagating quantum front that propagates ballistically in the lower density regions. In the case of fermions on the lattice (or spin chains), this front exhibits a staircase-like fine structure in density \cite{hunyadi2004dynamic,gobert2005real,moriya2019exact}. Interestingly, this phenomenon was related \cite{eisler2013full} to the statistics of eigenvalues at the edge of the GUE \cite{forrester1993spectrum,tracy1994level,tracy1999airy}, where the FCS takes the form of a Fredholm determinant involving the so-called Airy kernel. For the purpose of this article, it is useful to recall the main results of these works \cite{hunyadi2004dynamic,gobert2005real,moriya2019exact,eisler2013full}. 

\begin{figure}[ht]
\centering
\makebox[\textwidth][c]{\includegraphics[width=.9\linewidth]{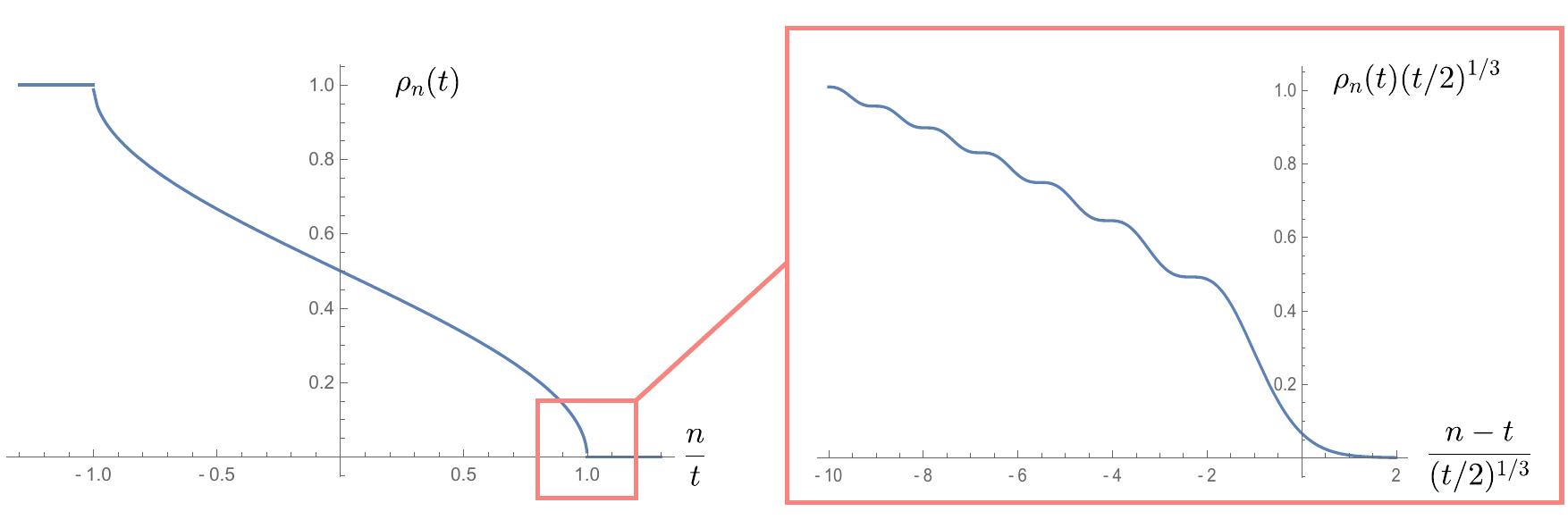}}\captionsetup{width=.85\linewidth}
\caption{Illustration of the density $\rho_n(t)=C_{n,n}(t)$. On the left panel \cite{antal1999transport}, we plot the large time scaling form of the density \eqref{eisler_density}. We observe the propagation of a front in the density. On the right panel, we show the density rescaled around the front which follows Eq. \eqref{eisler_result}. There is a staircase structure at the edge of the front \cite{hunyadi2004dynamic}.}
\label{fig:eisler}
\end{figure}

The authors of \cite{eisler2013full} consider spinless free fermions on a lattice. The corresponding discrete Hamiltonian is given by:
\al{\label{discrete_h}\hat H_d=-\frac{1}{2}\sum\limits_{m=-\infty}^{+\infty} (c^\dagger_m c_{m+1}+c^\dagger_{m+1} c_{m}),}
where $c^\dagger_m$ and $c_m$ are respectively the fermionic creation and annihilation operators at site $m$. Initially, the system is prepared in a step like condition filling every sites of the left half of the system 
\be\label{initialcond1}
\braket{c_m^\dagger c_n}=
\begin{cases}
    \delta_{n,m} \quad n,m  \leq 0\\
    0 \quad \text{else}
\end{cases}.
\ee
Now we are interested in the time-dependent correlation, i.e. $C_{m,n}(t)=\braket{c_m^\dagger (t) c_n(t)}$ where $c_n(t)$ is the time evolved fermionic annihilation operator in the Heisenberg picture. More precisely, a propagating front moving from left to right emerges, and we are interested in the correlation around this front. It can be shown that the density in the large time limit takes a scaling form (see Fig. \ref{fig:eisler})
\al{\label{eisler_density}\rho_n(t)\simeq \begin{cases}
    1 \hspace*{2.15cm}\mbox{ if } x/t<-1\\
    \arccos(x/t)/\pi \hspace*{0.1cm} \mbox{ if } -1<x/t<1\\
    0 \hspace*{2.2cm}\mbox{ if } 1<x/t
\end{cases}. }
The correlations can be computed in the large time limit with a saddle point approximation. However, near the quantum front, the saddle point is by definition  located at the maximum of the velocity $v(k)=\epsilon'(k)$ (the dispersion relation is $\epsilon(k)=-\cos(k)$) that is the saddle point is at $v(k^*)=\sin(k^*)=1$. As a consequence, the expansion around the saddle point has a vanishing second order term at the quantum front. Hence, the resulting third order term $\sim k^3$ leads to the emergence of the Airy kernel \cite{eisler2013full}
\be\label{eisler_result}
C_{m,n}(t)\simeq i^{n-m} (\frac{2}{t})^{1/3}K_{\Ai}\left(\frac{m-t}{(\frac{t}{2})^{1/3}},\frac{n-t}{(\frac{t}{2})^{1/3}}\right),
\ee
where the Airy kernel defined as
\al{\label{airy_kernel}K_{\Ai}(x,x')=\frac{\Ai(x) \Ai'(x')-\Ai(x') \Ai'(x)}{x-x'},}
and $\Ai(x)$ refers to the Airy function, i.e. the solution of the equation 
\al{ y''-xy=0,} with the initial conditions
\al{y(0)=\frac{1}{3^{2/3}}\Gamma(\frac{2}{3}), \mbox{ and } y'(0)=-\frac{1}{3^{1/3}}\Gamma(\frac{1}{3}).}
The Airy kernel was initially found at the edge of  the Gaussian Unitary Ensemble (GUE) \cite{forrester1993spectrum} and then led to the so called Tracy-Widom distribution \cite{tracy1994level}, i.e. the distribution of the largest eigenvalue of the GUE. We see that $t^{1/3}$ emerges as a characteristic length. Note that the Wick theorem implies that the phase factor $i^{n-m}$ in front of the kernel is irrelevant for the density-density correlations.\par

Now we compare this with fermions in the continuum whose dynamics where already analysed in \cite{bettelheim2012quantum,dean2019nonequilibrium,detbp}. We will focus on a bipartition protocol as this was the case in \cite{Collura_2014}. The one particle Hamiltonian is given by
\al{\hat H=-\frac{1}{2}\partial_x^2.}
In that case, the dispersion relation is $\epsilon(k)=\frac{k^2}{2}$, therefore the velocity $v(k)=k$ has no extremum. This implies that we will never witness the emergence of the Airy kernel as it is the case for fermions in the lattice \cite{eisler2013full}. Despite this, we propose a quench protocol for fermions in the continuum where the Airy kernel Eq. \eqref{airy_kernel} arises at the edge of the quantum front. Note that a similar phenomenon was already observed in a different context in \cite{bettelheim2012quantum}. However, the phenomenon we observed is distinct, and the mathematical approach we took has given rise to new questions. The main trick is to change the initial condition. Instead of the domain wall initial condition as in \cite{eisler2013full, Collura_2014}, we prepare the system in the ground state of the inverse power-law potential $V(x) = \frac{c}{|x|^\gamma}$, such that only the left half-space is filled up to the Fermi momentum $p_F$. Then, at the initial time, we turn off the potential, allowing the fermions to freely propagate. From here, the statistics at the edge of the quantum front can be related to the Wigner function in the initial state. Therefore, following \cite{dean2018wigner}, we examine the initial state Wigner function corresponding to the ground state of the inverse power-law potential, and we demonstrate the emergence of the Airy kernel Eq. \eqref{airy_kernel} at the edge of the quantum front. Additionally, this comes with anomalous diffusion around the quantum front with a continuum of generalized diffusion coefficient.\par

\section{Main results}
\subsection{The Wigner function and the inverse power law potential}\hfill\\
The first result of this article is the careful study of the ground state Wigner function $W(x,p)$ \cite{wigner1997quantum} (defined below in \eqref{wigner_func_def}) in the repulsive inverse power law potential $V(x)=\frac{c}{|x|^\gamma}$ in the large chemical potential $\mu$ limit or in the small $\hbar$ limit.\par
\begin{figure}[H]
    \centering
    \includegraphics[width=.7\linewidth]{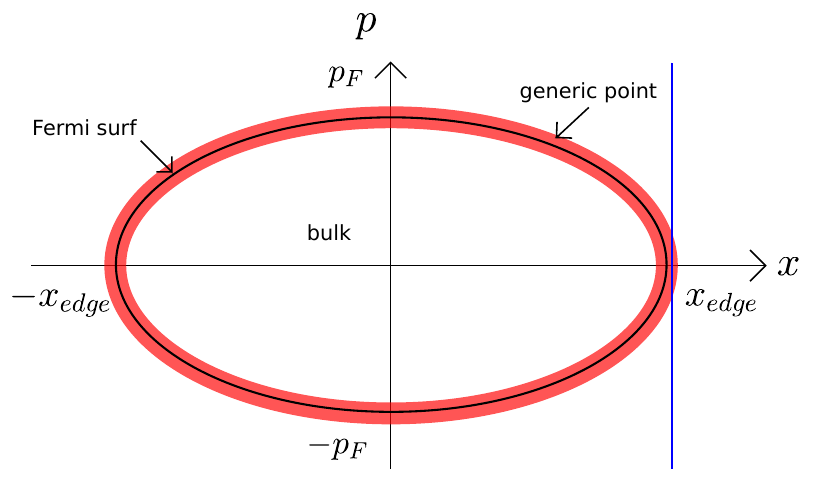}\captionsetup{width=.9\linewidth}
    \caption{Representation of the Fermi surf for the harmonic potential. We also represent in red, the width of the Wigner function in the semiclassical limit around the Fermi surf. Additionally, the vertical line in blue is the integral contour of the formula \eqref{kernel_wigner} which leads to the recovery of the Airy kernel at the edge of the gas \eqref{airy_kernel3}.}
    \label{fig:harmonic_fermi}
\end{figure}

We first recall already known properties of the Wigner function in the case of a smooth potential. It is already known \cite{castin2007basic,dean2019noninteracting} that in both the limits of large $\mu$ and small $\hbar$, for non interacting fermions in a smooth potential, the Wigner function has the limit
\be\label{classical_wigner}
W(x,p) \simeq \frac{1}{2\pi\hbar}\Theta(\mu-H(x,p)),
\ee
where $\Theta(x)$ is the Heaviside theta function, $\mu=\frac{p_F^2}{2m}$ is the chemical potential, and $H(x,p)=\frac{p^2}{2m}+V(x)$ is the classical one particle Hamiltonian of a particle in a smooth potential $V(x)$. The surface defined by
\be\label{fermi_surf1}
H(x_s,p_s)=\mu,
\ee
is called the Fermi surf, this is illustrated in the case of a harmonic potential in Fig. \ref{fig:harmonic_fermi}.
Additionally to that, it is known that for any generic point  close to the Fermi surf, the Wigner function has the subdominant universal behaviour in the large $\mu$ limit \cite{dean2018wigner}:
\al{\label{Airyscaling}
& W(x,p) \simeq \frac{\mathcal W (a)}{2\pi\hbar}=\frac{1}{2\pi \hbar}\int_{2^{2/3}a}^\infty dz \Ai(z),\\
& a= \frac{H(x,p)-\mu}{e}, \quad e=(\frac{\hbar^2}{2}(p_s^2 V''(x_s)+(V'(x_s))^2)^{1/3},\nn
}
where $\Ai$ is the Airy function.  We say this scaling is universal in the sense that it is valid for any smooth potential as long as $e\neq 0$. Note that 
\be
\mathcal W (a) \simeq \begin{cases}
    1 \quad \quad \quad \quad \quad \quad \quad \quad \quad \quad \quad \quad \mbox{ if } a \to -\infty\\
    (8\pi)^{-1/2} a^{-3/4} \exp\left(-\frac{4}{3} a^{3/2}\right)\quad  \mbox{if } a \to +\infty
\end{cases}.
\ee
This provides a smooth matching between the two behaviours of the Wigner function on each sides of the Fermi surf in \eqref{classical_wigner}. Furthermore, let us define the correlation function as
\al{\label{corr_def}C(x,x',t)=\braket{c_{x}^\dagger(t)c_{x'}(t)},}
with $c_{x}^\dagger(t)$ the time evolved fermionic creation operator in the Heisenberg representation. Then, using the formula from \cite{olivier2010airy} 
 \be\label{book_formula}
 \int \frac{dk}{\pi} e^{-ik(p-q)}\Ai(k^2+p+q)=2^{2/3}\Ai(2^{1/3} p) \Ai(2^{1/3} q) ,
\ee
together with some changes of variables, and the formula \eqref{kernel_wigner} relating the correlation function to the Wigner function, one can show \cite{dean2018wigner} that the correlation function at the edges of the gas is related to the Airy kernel \eqref{airy_kernel} for any smooth potential 
\be\label{airy_kernel3}
C(x,x')\underset{N\to\infty}{\simeq} K_{\Ai}(\frac{x-x_{edge}}{w_N},\frac{x'-x_{edge}}{w_N}), \quad w_N=\left( \frac{\hbar^2}{2mV'(x_{edge})}\right)^{1/3},
\ee
where $x_{edge}$ is the edge of the particle density defined by $V(x_{edge})=\mu$, not to be confused with the Fermi surf $(x_s,p_s)$ (see Fig. \ref{fig:harmonic_fermi}).\par

The Airy scaling \eqref{Airyscaling} is valid in a smooth potential only  close to any generic point of the Fermi surf, i.e. for $V(x_s)\simeq \mu$ and $p_s\simeq \mu^{1/2}$. The case of the harmonic potential was studied in detail and the Airy scaling \eqref{Airyscaling} was shown to be valid close to any point of the Fermi surf, including non generic points \cite{dean2018wigner}.\par
Now, we consider the particular case of the repulsive inverse power law potential. We look at the complete Fermi surf which includes non generic points and we ask the question where the Airy scaling \eqref{Airyscaling} is valid. For this purpose, we split the Fermi surf into two different regions, near the gas edge $x_{edge}$ and away from it.\par
\hspace{1cm}

\begin{figure}[ht]
    \centering\includegraphics[width=0.85\textwidth]{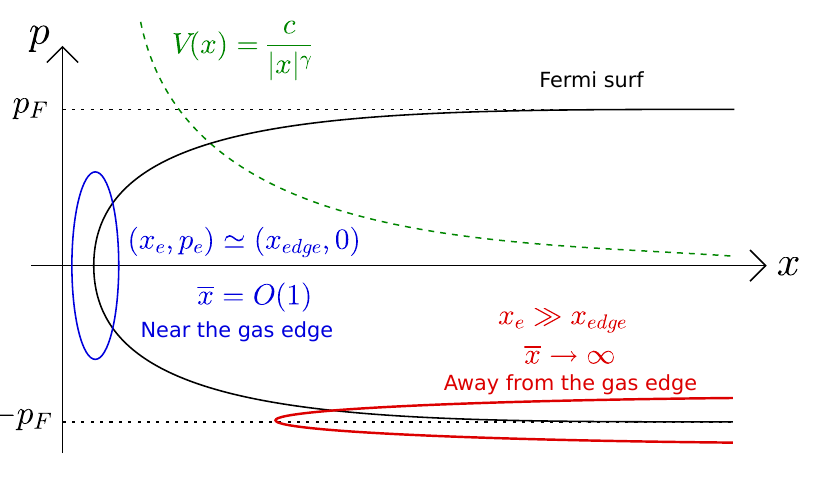}\captionsetup{width=1\linewidth}
    \caption{Illustration of the Fermi surf of the Wigner function for the ground state of the repulsive inverse power law potential $V(x)=\frac{c}{|x|^\gamma}$. We specify the two regions of the Fermi surf defined in terms of $\bar{x}$ of order one or taken large where $\bar{x}$ is defined in \eqref{nearscaling}.}
    \label{fig:wigner_regime}
\end{figure}

{\bf Near the gas edge.} We want to look at the Wigner function near the gas edge that is where the particle density vanishes. The gas edge is defined by $p_s=0$, and $x_s=x_{edge}=(\frac{c}{\mu})^{1/\gamma}$ (see Fig. \ref{fig:wigner_regime}). We define the rescaling of the Fermi surf as 
\al{\label{nearscaling}&x_s=\bar x x_{edge},\\
&p_s=\sqrt{2m\mu(1-\bar x^{-\gamma})},\nn}
where $\bar{x}\in [1,+\infty[$. Then we consider the different limits (large $\mu$ or small $\hbar$) with $\bar x$ fixed. 
In that case, we showed (see Section \ref{art41}) that the Airy scaling \eqref{Airyscaling} is valid under the condition
\be\label{nearcond}
\frac{\hbar^2 \mu^{\frac{2-\gamma}{\gamma}}}{m c^\frac{2}{\gamma}}\ll 1.
\ee
Hence, we find the Airy scaling Eq. \eqref{Airyscaling} in two different limits, for $\mu \to \infty, \quad \gamma >2$, or for $ \hbar \to 0$ with any $\gamma>0$. In other words, the limit $\hbar \to 0$ yields the Airy scaling for any $\gamma$. Notice that if we set $c=\frac{\hbar^2(\nu^2-\frac{1}{4})}{2m}$ (this would be the case if we proceed to the spherical harmonic decomposition of fermions in $d$ dimensions), the previous condition becomes
\be
\frac{1}{m (\nu^2-\frac{1}{4}))^\frac{2}{\gamma}}(\frac{\hbar^2}{ \mu})^{\frac{\gamma-2}{\gamma}}\ll 1.
\ee
Thus, for small $c\sim\hbar^2$ barrier, the small $\hbar$ and the large $\mu$ limit are equivalent and the Airy scaling is valid only if $\gamma>2$, which agrees with \cite{dean2016noninteracting,lacroix2018non}.\par
\hspace{1cm}

{\bf Away from the gas edge.} We look at the Wigner function for $x_{edge} \ll x_s$ and  $p_s \simeq \sqrt{2m\mu}$ (see Fig. \ref{fig:wigner_regime}) or equivalently $\bar x \to \infty$ in \eqref{nearscaling}. 
In the limits of interest, we showed (see Section \ref{art41}) that the the Airy scaling is correct only if the following condition is satisfied
\be\label{expcond5}
\frac{\hbar \sqrt{\mu} x_s^{\gamma-1}}{\sqrt{m}c}\ll 1.
\ee
Let us make a few remarks on this condition:
\begin{itemize}
    \item[$\bullet$] The large $\mu$ limit does not satisfy the condition \eqref{expcond5}, hence, there is no Airy scaling at large $\mu$, and large $x_s$ for both of those limits.
    \item[$\bullet$] In the case of the small $c=\frac{\hbar^2(\nu^2-\frac{1}{4})}{2m}$ barrier, again, the large $\mu$ and small $\hbar$ limit play the same role and there is no Airy scaling at large $x_s$.  
    \item[$\bullet$] There is a change of behavior at $\gamma=1$. For $\gamma>1$ the Airy scaling is recovered for small $\hbar$ as long as $x_s$ is small enough to verify Eq. \eqref{expcond5}. For $\gamma\leq 1$ the Airy scaling is recovered for any $x_s$. 
    \item[$\bullet$] The question that remains open is how the Wigner function behaves near the Fermi surf when the condition Eq. \eqref{expcond5} is broken as $x_s$ is taken large.
\end{itemize}

The limit leading to the Airy scaling are summarized in the table \ref{wigner_regime_t}.\par
\begin{table}[H]
\centering
\begin{tabular}{|c|c|c|}
\hline 
Limit & {\blue Near the gas edge} & {\red Away from the gas edge}  \\
\hline
$\mu \to +\infty$ & $\gamma>2$  & $\emptyset$ \\
\hline
$\hbar \to 0$ &  $\forall \gamma$    & $\forall x \text{ if } \gamma <1 \text{ or if }  x^{\gamma-1}\ll \frac{\sqrt{m}c}{\hbar \sqrt{\mu}}$  \\
\hline
\end{tabular}
\captionsetup{width=\linewidth}
\caption{Summary of the conditions under which the Wigner function exhibits the Airy scaling \eqref{Airyscaling} along the Fermi surf (see Fig. \ref{fig:wigner_regime}) for the ground state in the inverse power law potential.}\label{wigner_regime_t}
\end{table}
\subsection{Airy kernel at the edge of the propagating front}\hfill\\
The second result of this article is stated as follows. A gas of fermions is prepared in the ground state of the repulsive inverse power law potential, such that only the left half space is filled up to momentum $p_F$ (or chemical potential $\mu=\frac{p_F^2}{2m}$) with small $\hbar$. Then we quench the potential such that the fermions evolve freely and a front propagates with a positive velocity. Our result is the emergence of the Airy kernel \eqref{airy_kernel} at the edge of the propagating front. 
More precisely, close to the edge $x_{edge}(t)$ of the propagating front, the correlation function takes the form
\begin{align}\label{result2}
&C(x,x',t)\simeq -\frac{1}{w(t)} e^{-\frac{i}{\hbar}( p(t)(x-x')+\frac{(x-x_{edge}(t))^2}{2t}-\frac{(x'-x_{edge}(t))^2}{2t})}\nn\\
& \quad \quad \quad \quad \quad \quad \times K_{\Ai}(\frac{x-x_{edge}(t)}{w(t)},\frac{x'-x_{edge}(t)}{w(t)}),\\
&w(t)=t(\frac{\hbar^2 |p_s''(x(t))|}{2})^{1/3}=(\hbar^2 \frac{\gamma +1}{2}(\frac{p_F}{\gamma c})^{1/(\gamma+1)})^{1/3} t^{\frac{2\gamma+1}{3(\gamma+1)}},\nn
\end{align}
where $x_{edge}(t)$ is defined in \eqref{xedge}, and $x(t)$, $p(t)$ are defined in \eqref{Quantum_front}. Let us make a few remarks:
\begin{itemize}
    \item At large time $w(t)\sim t^{\frac{2\gamma+1}{3(\gamma+1)}}$. For $\gamma\to 0$ we recover the $\sim t^{1/3}$ from \eqref{eisler_result} and for $\gamma<1$ the process is subdiffusive. For $\gamma=1$ the propagation is diffusive and it becomes super diffusive for $\gamma>1$ up to $\sim t^{2/3}$ when $\gamma\to \infty$.
    \item Notice that the phase in front of the kernel is not relevant for the computation of the density density correlation. Indeed, if we note the correlations with a phase $C_f(x,y)=e^{f(x)-f(y)}C(x,y)$, where $f(x)$ an arbitrary function of $x$, the density density correlation is given by \al{\mathcal{R}(x_1,...,x_n)=\det\limits_{1\leq i,j \leq n} |C_f(x_i,x_j)]=\det\limits_{1\leq i,j \leq n} |C(x_i,x_j)|.} However in the case of non-equilibrium dynamics of fermions, observables like the current are affected. Indeed, we recall that the particle current is a functional of the kernel as
    \al{J\left[C\right](x,t)=\frac{1}{2i}(\partial_y-\partial_x)C(x,y,t)|_{x=y}.}
    This implies that the phase modifies the current as
    \al{J\left[C_f\right](x,t)=J\left[C\right](x,t)+if'(x)\rho(x,t),}
    where $\rho(x,t)=C(x,x,t)$. In our specific case, because the Airy kernel alone produces no current this yields
    \al{J(x,t)=\left(p(t)+\frac{x-x_{edge}(t)}{t}\right)\rho_{\Ai}\left(\frac{x-x_{edge}(t)}{w(t)}\right),}
    where we used the notation $\rho_{\Ai}(x)=K_{\Ai}(x,x)$. Notice that the first term can be interpreted as a consequence of the motion of the quantum front with velocity $p(t)$.
    \item Here, the Airy kernel at the quantum front is already present at initial time. This does not have to be the case, for example if the initial potential behaves like the power law potential for large $x$ but has a derivative that cancels precisely at the edge of the gas ($V'(x_{edge})=0$). Then, the Wigner function at initial time exhibits the Airy scaling \eqref{Airyscaling} only far away from edge of the gas, such that we expect the Airy kernel to appear only in the large time limit.
    \item In \cite{dean2019nonequilibrium}, a small $\hbar$ analysis of the Wigner function propagation within a potential is done. The conclusion is that if present at initial time, the Airy scaling \eqref{Airyscaling} would persist at short time (up to some rescaling of the width $e$).
    \item The Airy kernel \eqref{result2} is valid only if the initial time Wigner function exhibits the Airy scaling at position $x(t)$, that is if $x(t)$ verifies the condition  \eqref{expcond5} from the first section.
    \item This quench set up appears as a possible experimental protocol to measure fine features of the Wigner function in an equilibrium state. It allows to probe the Airy scaling of the ground state Wigner function when $|x_s| \to \infty$. This is because the Airy scaling behaviour of the pre-quench Wigner function is related to the Airy kernel emerging at the propagating front in the large time limit.
\end{itemize} 
The rest of the article is organised as follows. The section \ref{art41} is dedicated to the study of the initial condition, that is the ground state properties of non interacting fermions in the inverse power law potential $V(x)=\frac{c}{|x|^\gamma}$. More precisely, our goal is to explore the properties of the semiclassical limit of the Wigner function and to prove the result given in the table \ref{wigner_regime_t}.

In the section \ref{art42}, we introduce a quench protocol where noninteracting fermions are prepared in the ground state of the inverse power law potential $V(x)=\frac{c}{|x|^\gamma}$, occupying only the left half of the system. This way the initial condition corresponds to the Wigner function studied in the precedent section \ref{art41}. Then, the potential is quenched to zero, allowing the fermions to evolve freely. Consequently, a quantum front (or the edge of the gas) propagates from left to right. In section \ref{art42}, as a consequence of section \ref{art41}, we give a derivation for \eqref{result2}, that is the emergence of the Airy kernel around the quantum front. The article is closed by a concluding section \ref{Conclusion}. Finally, we have two appendices. In Appendix 
\ref{app:gamma2} we proceed to the analysis of the Wigner function in the particular case of the inverse squared potential. In Appendix \ref{app:inversepowerlaw} we provide the analysis of the Wigner function for the general repulsive inverse power law potential. 

\section{Airy statistics and the Wigner function}\label{art41}

\subsection{Wigner function}\hfill\\
In quantum mechanics, the position probability density function is given by the squared modulus of the wave function, $\rho(x) = |\psi(x)|^2$, where $\rho$ represents the spatial density. Similarly, the momentum density function is related to the squared modulus of the Fourier transform of the wave function, $\hat{\rho}(p) = |\hat{\psi}(p)|^2$, where $\hat{\psi}$ is the Fourier transform of $\psi$, and $\hat{\rho}$ represents the momentum density. However, due to the uncertainty principle, a well-defined joint probability density function (JPDF) in the position-momentum phase space $(x, p)$ cannot be defined.\par

To address this issue, Wigner introduced the concept of the Wigner function in 1932 \cite{wigner1997quantum}. The Wigner function provides an attempt at describing the phase-space properties of quantum systems. For a wave function of $N$ fermions in one dimension, the Wigner function is defined as follows:
\be\label{wigner_func_def}
W(x,p)=\int_{-\infty}^{+\infty} \frac{dy}{2\pi\hbar} \prod\limits_{i=2}^N dx_i e^{\frac{ipy}{\hbar}}\psi^*(x+\frac{y}{2},x_2,...,x_N)\psi(x-\frac{y}{2},x_2,...,x_N).
\ee
From this, integrating over momentum, or position space, yields the spatial and momentum particle densities 

\al{\label{densitywdef}\int_{-\infty}^{+\infty} dp  W(x,p)&= \rho_N (x),\quad \int_{-\infty}^{+\infty} dx W(x,p)=  \hat \rho_N (p),\nn\\ &\iint_{-\infty}^{+\infty} dx dp W(x,p)= N.}
We used the definition
\be 
\rho_N (x)=\braket{\sum\limits_{i=1}^N \delta(x-x_i)},\quad \hat \rho_N (p)=\braket{\sum\limits_{i=1}^N \delta(p-p_i)},
\ee
where $\braket{...}$ is the average over the quantum state of wave function $\psi$. Notice that the two densities are well normalised $\int dx \rho_N(x)=\int dp \hat \rho_N(p)=N$. However, the Wigner function does not represent a probability distribution as it can take negative values, although it remains a real-valued function.\par

In the case of noninteracting fermions, it can be related to the correlation function through the Weyl transform
\be\label{wigner_kernel}
W(x,p)=\int_{-\infty}^{+\infty} \frac{dy}{2\pi \hbar} e^{\frac{ipy}{\hbar}}C(x+\frac{y}{2},x-\frac{y}{2}).
\ee
Together with its inversion formula
\be\label{kernel_wigner}
C(x,x')=\int_{-\infty}^{+\infty} dp e^{-i\frac{p}{\hbar}(x-x')} W(\frac{x+x'}{2},p).
\ee
Now we want to delve into the semiclassical behaviour of the Wigner function.\par

First, let us provide a detailed justification for the classical limit \eqref{classical_wigner} of the Wigner function. By doing so, we will also refine this limit form and obtain the semiclassical Airy scaling \eqref{Airyscaling} around the generic points.

\subsection{Short time expansion of the propagator and the semiclassical Limit}\label{sec_shorttime}\hfill\\
In this subsection, we recall the derivation from \cite{dean2018wigner} which yields the semiclassical limits \eqref{classical_wigner} and \eqref{Airyscaling}. This derivation was given in the large $\mu$ limit close to a generic point of the Fermi surf. Here we restore $\hbar$ and explain why it yields the same limit if $\hbar$ is small. Then we will give details on the difference between both of those limits in the specific case of the inverse power law potential. We will discuss the behaviour of the Wigner function close to the complete Fermi surf, both at generic points and away from such points.
 \hspace{1cm}
 
{\bf Imaginary Time Propagator.} We consider a system of noninteracting trapped fermions with a single-particle Hamiltonian $\hat{H}$, wave functions, and eigenenergy family $\{\phi_k, \epsilon_k\}$ (with $k=0,1,2,...$). The system is in the ground state with chemical potential $\mu$ or equivalently a finite number of $N$ fermions. We note $C_\mu$ the function
\be
C_\mu(x,x')=\sum\limits_{k=0}^\infty \phi_k^*(x)\phi_k(x') \Theta(\mu-\epsilon_k).
\ee
Note that in the case of a non-trapping potential the family $ \phi_k, \epsilon_k $ have a continuously indexed component, the previous sum turns into an integral, and for a given chemical potential $\mu$ the number of fermions is infinite. Our method relies on the imaginary quantum propagator
\be\label{prop_def}
G(x,x',t)=\braket{x'|e^{-\frac{\hat H t}{\hbar}}|x}=\frac{t}{\hbar}\int_0^\infty d\mu e^{-\frac{\mu t}{\hbar}} C_\mu(x,x').
\ee
The last formula can be inverted to give the kernel as
\be\label{kernel_prop}
C_\mu(x,x') = \int_{\Gamma}\frac{dt}{2\pi i t} e^{\frac{\mu t}{\hbar}}G(x,x',t),
\ee
where $\Gamma$ is the Bromwich contour in the complex plane. Hence putting together Eq. \eqref{wigner_kernel}, and \eqref{kernel_prop}, yields the expression for the Wigner function in the ground state of chemical potential $\mu$
\be\label{wigner_prop}
W(x,p)=\frac{1}{2\pi\hbar} \int_{C}\frac{dt}{2\pi i t} e^{\frac{\mu t}{\hbar}} \int_{-\infty}^{+\infty} dy e^{\frac{ipy}{\hbar}} G(x+\frac{y}{2},x-\frac{y}{2},t).
\ee
The propagator obeys the following equation
\be\label{propagator_eq}
\hbar \partial_t G(x,x',t)=\left(\frac{\hbar^2}{2m}\partial_x^2 -V(x)\right)G(x,x',t),\quad G(x,x',0)=\delta(x-x').
\ee
Now, using the Feynamnn-Kac formula, the solution of this equation can be written as a path integral \cite{oksendal2003stochastic}. This can then be written in term of a Brownian bridge \cite{dean2016noninteracting}
\be\label{propagator}
G(x+\frac{y}{2},x-\frac{y}{2},t)= \sqrt{\frac{m}{2\pi \hbar t}}e^{-\frac{m y^2}{2\hbar t}} \langle \exp(-\frac{t}{\hbar}\int_0^1 du V(x + y(\frac{1}{2}-u)+\sqrt{\frac{\hbar t}{m}}B_u)) \rangle_B.
\ee
Here $B$ is a Brownian bridge that is a Gaussian process on $u\in[0,1]$ with zero mean and correlation $\braket{B_uB_{u'}}_B={\rm min}(u,u')-uu'$, hence $B_0=B_1=0$. From here, the argument is that in the semiclassical limit (where $\mu$ is large), the Wigner function, given by Eq. \eqref{wigner_prop}, will be dominated by short-time contributions. Therefore, we proceed with the short-time expansion of the propagator. There is a saddle point at $y=itp$, so we make the change of variable $y=\frac{itp}{m} + \sqrt{\frac{\hbar t}{m}} \tilde{y}$ which yields (appendix of \cite{dean2018wigner})
{\small \begin{align}\label{toexpand}
& W(x,p)=\frac{1}{2\pi \hbar}\int_C \frac{dt}{2\pi i t}e^{(\mu -\frac{p^2}{2m}-V(x))\frac{t}{\hbar} +S(x,p)}\\
& S(x,p)= \ln  \langle \exp\bigg(-\frac{t}{\hbar}\int_0^1 du \left[ V(x + \frac{itp}{m} + \sqrt{\frac{\hbar t}{m}} \tilde{y}(\frac{1}{2}-u)+\sqrt{\frac{\hbar t}{m}}B_u) - V(x)\right] \bigg) \rangle_{B,\tilde{y}},\nn
\end{align} }%
with $\braket{...}_{B,\tilde{y}}=\int \frac{d\tilde{y}}{\sqrt{2\pi}}e^{-\frac{\tilde{y}^2}{2}}\braket{...}_{B}$.\par
 \hspace{1cm}
 
{\bf Recovering the classical limit.} Now, our goal is to justify the classical limit \eqref{classical_wigner}. In order to do so, we expand $V$ in \eqref{toexpand} for $(x,p)$ close to $(x_s,p_s)$ at small time $t$ and we stop to the zeroth order in time. The two potentials $V$ in \eqref{toexpand} cancel, which yields $S(x,p)\simeq 0$, and therefore
\be
W(x,p)=\frac{1}{2\pi \hbar}\int_C \frac{dt}{2\pi i t}e^{(\mu -\frac{p^2}{2m}-V(x))\frac{t}{\hbar}}=\frac{1}{2\pi\hbar}\Theta(\mu-H(x,p)).
\ee
This is the result of Eq. \eqref{classical_wigner}. The only missing part in our derivation is a justification for the short time expansion. In the large chemical potential $\mu$ limit, this expansion was justified in \cite{dean2018wigner}.\par
 \hspace{1cm}
 
{\bf The semiclassical limit.} Now, we want to push this expansion to higher orders. We choose a point close to the edge, $(x,p)\simeq (x_s, p_s)$, and we expand $V$ in \eqref{toexpand} around $x_s$. The expansion is achieved in detail in the Appendix \ref{app:inversepowerlaw} considering the fluctuations of the Brownian bridge $B_u$. The expansion is found to be valid under the conditions given in the Appendix in Eqs. \eqref{expcond2}, and \eqref{expcond3}. For now let us assume that those conditions are satisfied. The expansion yields the following result

\al{\label{airy_scale_concle}
W(x,p) &\underset{N\to\infty}{\simeq} \frac{1}{2\pi \hbar}\int_C \frac{dt}{2\pi i t}e^{(\mu-H(x,p))\frac{t}{\hbar} + \frac{t^3}{24 \hbar }(p_s^2 V''(x_s)+V'(x_s)^2)},\\
&  =\frac{1}{2\pi \hbar}\int_C \frac{d\tau}{2\pi i \tau}e^{-2^{2/3} a\tau + \frac{\tau^3}{3}}=\frac{\mathcal W (a)}{2\pi \hbar},\nn
}
where $a$ was defined in Eq. \eqref{Airyscaling}. This is the Airy scaling discussed above in Eq. \eqref{Airyscaling}. The authors of \cite{dean2018wigner} then argued that in smooth confining potentials, close to a generic point of the Fermi surf, the conditions Eqs. \eqref{expcond2} and \eqref{expcond3} are verified in the large $\mu \to \infty$ limit, which allows to recover the universal scaling Eq. \eqref{Airyscaling} and \eqref{airy_kernel3}. From now on, the conditions Eqs. \eqref{expcond2} and \eqref{expcond3} will be essential if one wants to know whether the Wigner functions follows the Airy scaling \eqref{Airyscaling} in a specific region of the Fermi surf.\par

While the Airy scaling \eqref{Airyscaling} is generally valid for generic points, it has been found to be absent at certain points of specific potentials. For instance, when examining the Wigner function of potentials $V(x)\sim x^{2n}$ with $2\leq n$, it is clear that $e$ defined in \eqref{Airyscaling} becomes equal to zero at $x_s=0$, such that, the Airy scaling is not verified. This phenomenon is discussed in \cite{le2018multicritical}, where the authors obtained the momentum coordinate kernel at the edge of the momentum density and found it to differ from the Airy kernel. Another example is provided by the inverse power law potential $V(x)=\frac{\hbar^2 (\nu^2-\frac{1}{4})}{2 m|x|^\gamma}$, for which the kernel at the edge of the density was determined in \cite{lacroix2018non}. This analysis shows the emergence of the Airy kernel for large $\mu$, if $\gamma>2$. Knowing this, we aim to extend the analysis to the full Wigner function and investigate the conditions under which the Wigner function exhibits the Airy scaling \eqref{Airyscaling} for the inverse power law potential.\par
 \hspace{1cm}
 
{\bf Remark.} It is interesting to note that the short time expansion of the propagator can be obtained from an expansion around the classical path, as demonstrated in \cite{makri1989exponential}. In the case of an inverse power law potential $V(x) = \frac{c}{x^\gamma}$, there are two such paths: a direct path and an indirect path with a turning point \cite{bhagwat1989new, lolle1991improved}. However, when using the imaginary time propagator in the small $\hbar$ limit, the contribution from the indirect path is subleading such that the two different methods yield identical results.\par

\subsection{The Inverse Repulsive Power Law Potential \texorpdfstring{$V(x)=\frac{c}{|x|^\gamma}$}{Lg}} \hfill\\
Here for purpose that will become clear in the section \ref{art42} we want to extend the previous analysis to the special case of the non-confining potential $V(x)=\frac{c}{|x|^\gamma}$ with $c>0$. We show that for this potential, the Airy scaling Eq. \eqref{Airyscaling} in the large $\mu$ limit is valid only under tight conditions, that is for $\gamma>2$ and close to the edge of the gas (or equivalently on the part of the Fermi surf Eq. \eqref{fermi_surf1} where $p_s\simeq0$).  Our main result relies in the fact that the Airy scaling remains valid on a wider part of the Fermi surf in the small $\hbar$ limit. More precisely as $\hbar \to 0$, the Airy scaling is valid on the complete Fermi surf if $\gamma \leq 1$, and only for $\hbar x_s^{\gamma-1}\ll 1$ if $\gamma>1$. Our results are summarized in the table \ref{wigner_regime_t} and figure \ref{fig:wigner_regime}.

In Ref. \cite{dean2018wigner}, the analysis was carried out with $\hbar=1$. Here, we reintroduce $\hbar$
, and we realize that the limit $\hbar \to 0$ leads to the Airy scaling \eqref{Airyscaling} on a wider part of the Fermi surf than in the large $\mu$ limit.

 \hspace{1cm}
 
{\bf The inverse square potential, $\gamma=2$.} 
Before examining the general inverse power law potential, we start with a particular case, the inverse squared potential. For $\gamma=2$, the eigenfunctions are known such that one can compute exactly the Wigner function. We know \cite{lacroix2018non} that the Airy kernel is recovered when $\nu\to\infty$, for fixed $c$, which is equivalent to taking the $\hbar \to 0$ limit, as  $\nu=\sqrt{\frac{1}{4}+\frac{2mc}{\hbar^2}}$. Therefore, we consider the Wigner function in the latter limit and prove that it exhibits the Airy scaling \eqref{Airyscaling}. In this case, the single-particle eigenfunctions $\phi_p$ form a continuously indexed family, and they are related to the Bessel functions through the following expression 
\be
\phi_p(x)=\sqrt{\frac{px}{\hbar}}J_{\nu}\left(\frac{px}{\hbar}\right),
\ee
with $\nu \simh \frac{\sqrt{2mc}}{\hbar}$. This leads to the following expression for the Wigner function \cite{de2021wigner}

\al{\label{gamatwowigner}
&W(x,p)=\frac{1}{2\pi \hbar^3}\int_0^{p_F} p'dp' \int_{-2x}^{2x} dy e^{\frac{i \nu y}{x_\alpha(p)}} \sqrt{(x+\frac{y}{2})(x-\frac{y}{2})}	 \\
&\quad \quad \quad \quad \times J_\nu\left(\nu\frac{x+y/2}{x_\alpha(p')}\right) J_\nu\left(\nu\frac{x-y/2}{x_\alpha(p')}\right),\nn
}
where $x_\alpha(p')=\frac{\sqrt{2mc}}{p'}$ is the turning point (i.e. the point where $\frac{p'^2}{2}=V(x_\alpha)$) for a wave of impulsion $p'$. Using the integral representation for the Bessel function $ J_\nu(x) = \frac{1}{2\pi} \int_{-\pi}^\pi d\tau e^{i(\nu \tau - x \sin(\tau))} $ in \eqref{gamatwowigner} leads to
{\small \al{\label{saddlepointform}
&W(x,p)=\frac{1}{(2\pi \hbar)^3}\int_0^{p_F} p'dp' \int_{-2x}^{2x} dy  \int_{-\pi}^\pi d\tau_1 d\tau_2\sqrt{x^2-\frac{y^2}{4}}  e^{i\nu F (\tau_1,\tau_2,p',y)}\nn\\
&F(\tau_1,\tau_2,p',y)=
\tau_1+\tau_2 - \frac{x+\frac{y}{2}}{x_\alpha(p')} \sin(\tau_1)- \frac{x-\frac{y}{2}}{x_\alpha(p')} \sin(\tau_2)+\frac{y}{x_\alpha(p)}.
}}%
In the small $\hbar$ limit, $\nu\to \infty$, hence we will evaluate the four dimensional integral Eq. \eqref{saddlepointform} using a saddle point method. We notice that the $y$ dependent integral converges towards a delta Dirac function in the small $\hbar$ limit. 
\al{ \int_{-2x}^{2x} dy \sqrt{x^2-\frac{y^2}{4}} e^{i\nu y \left(\frac{\sin(\tau_2)-\sin(\tau_2)}{2 x_\alpha(p')}+\frac{1}{ x_\alpha(p)}\right)}\underset{\nu \to \infty}{\simeq} \frac{1}{\nu}\delta\left(\frac{\sin(\tau_2)-\sin(\tau_2)}{2 x_\alpha(p')}+\frac{1}{ x_\alpha(p)}\right)}
When expanding the function  $F$ to second order, the delta Dirac part, that is the term proportional to $y$ constrain $\tau_1$ and $\tau_2$ to a one dimensional curve such that their second derivative contributions cancel out close to the saddle point. Therefore one has to push the expansion of $F$ to the third order. For a phase space point taken close to the Fermi surf $(x,p)\simeq (x_s,p_s)$, the third order terms lead to the Airy scaling limit Eq. \eqref{Airyscaling}. See Appendix \ref{app:gamma2} for details of this saddle point approximation. Additionally, we find that the saddle point approximation leading to the Airy scaling is valid under the following condition
\be\label{expcond1}
\frac{xp\hbar}{mc}\ll 1.
\ee
A few remarks are in order:
\begin{itemize}
    \item[$\bullet$] Notice that if the Airy scaling was obtained here in the limit $\hbar \to 0$, it would not be possible to obtain it solely from the large $\mu$ limit. 
    \item[$\bullet$] Typically $p$ is lower or close to $\sqrt{2m\mu}$ and the condition \eqref{expcond1} is not satisfied away from the potential that is for large $x\sim \hbar^{-1}$. 
\end{itemize}\par
 \hspace{1cm}
 
{\bf The inverse power law potential, $\gamma\geq 0$.} In order to generalise the result obtained for $\gamma=2$, we apply the short time propagator expansion method \eqref{sec_shorttime} to the inverse power law potential. Note that one could worry about the fact that the potential is not smooth anymore and exhibits a diverging singularity eventually invalidating the previous analysis. In fact, this is not a problem, as long as the singularity is repulsive, in this case the propagator from \eqref{propagator} is still well defined.\par
 \hspace{1cm}
 
{\bf Near the gas edge.} We want to look at the Wigner function near the gas edge that is where the particle density vanishes. The gas edge corresponds to the area of the Fermi surf around $p_s=0$, and $x_s=x_{edge}=(\frac{c}{\mu})^{1/\gamma}$ (see Fig. \ref{fig:wigner_regime}). The correct rescaling of the Fermi surf  is defined as
\al{\label{nearscaling1}&x_s=\bar x x_{edge},\\
&p_s=\sqrt{2m\mu(1-\bar x^{-\gamma})},\nn}
where $\bar{x}\in [1,+\infty[$. Then we consider the different limits (large $\mu$ or small $\hbar$) with $\bar x$ fixed. In that case, the conditions for the validity of the Airy scaling are given in Appendix \ref{app:inversepowerlaw}, Eqs. \eqref{expcond2}, and \eqref{expcond3}. Those conditions yield Eq. \eqref{omega1} which depends only on one dimensionless parameter $\omega=\frac{\hbar^2}{\mu^{\frac{\gamma-2}{\gamma}}c^{\frac{2}{\gamma}}}$. Finally, the expansion of the propagator is valid if this parameter is small that is if the single condition Eq. \eqref{nearcond} is satisfied. We recall that the short time expansion of the propagator has an interpretation in terms of classical paths. In the inverse power law potential there are two such paths, a direct path and an indirect path. Notice that the contribution of the indirect path can be neglected unless the condition Eq. \eqref{nearcond} is not respected. \par
\hspace{1cm}


{\bf Away from the gas edge.} We look at the Wigner function for $x_{edge} \ll x_s$ and  $p_s \simeq \sqrt{2m\mu}$ (see Fig. \ref{fig:wigner_regime}) or equivalently $\bar x \to \infty$ in \eqref{nearscaling1}. This time the conditions Eqs. \eqref{expcond2} and \eqref{expcond3} yield Eq. \eqref{expcond4}, such that, in the limits of interest, the Airy scaling is correct only if the condition \eqref{expcond5} is satisfied. This close the proof of the results \eqref{nearcond} and \eqref{expcond5}. Notice that the condition \eqref{expcond5} matches \eqref{expcond1} for $\gamma=2$.

\section{Airy kernel at the quantum front}\label{art42}

Now let us give details on the proposed quench protocol that will yield a propagating front with the Airy kernel \eqref{result2}.\par

{\bf Initial condition.} Instead of having a domain wall release like in \cite{eisler2013full} the system is prepared in the ground state of noninteracting fermions with the impenetrable barrier $V(x)=c/|x|^\gamma$, studied in Section \ref{art41}. This ground state is such that only the left half space is filled up to Fermi momentum $p_F$ or chemical potential $\mu=\frac{p_F^2}{2}$. 
The initial Wigner function $W_0(x,p)$ exhibits a  Fermi surf \eqref{fermi_surf1}. For the inverse power law potential (see Fig. \ref{fig:wigner_regime}), the Fermi surf $(x_s,p_s)$ at a given Fermi momentum $p_F$ follows the equaion
\al{
&\frac{p_s^2}{2}+\frac{c}{|x_s|^\gamma}=\frac{p_F^2}{2}.
}

Additionally near the Fermi surf, in the small $\hbar$ limit, the initial Wigner function $W_0(x,p)$ exhibits the Airy scaling \eqref{Airyscaling} which by taking the derivatives of \eqref{fermi_surf1}, takes the form
\al{\label{Airyscaling1}
&W_0(x,p)\simeq\frac{\mathcal{W}(\kappa(x) \delta{p}(x))}{2\pi \hbar},\nn\\
& \kappa(x) = (\frac{\hbar^2 p_s''(x)}{2})^{-1/3} \underset{x\to\infty}{\simeq} (\frac{2p_F |x|^{\gamma+2}}{\hbar^2 \gamma(\gamma+1)c})^\frac{1}{3},\\
& \delta{p(x)}=p-p_s(x).\nn
}
 \hspace{1cm}
 
\begin{figure}[ht]
\centering
\makebox[\textwidth][c]{\includegraphics[width=1.2\linewidth]{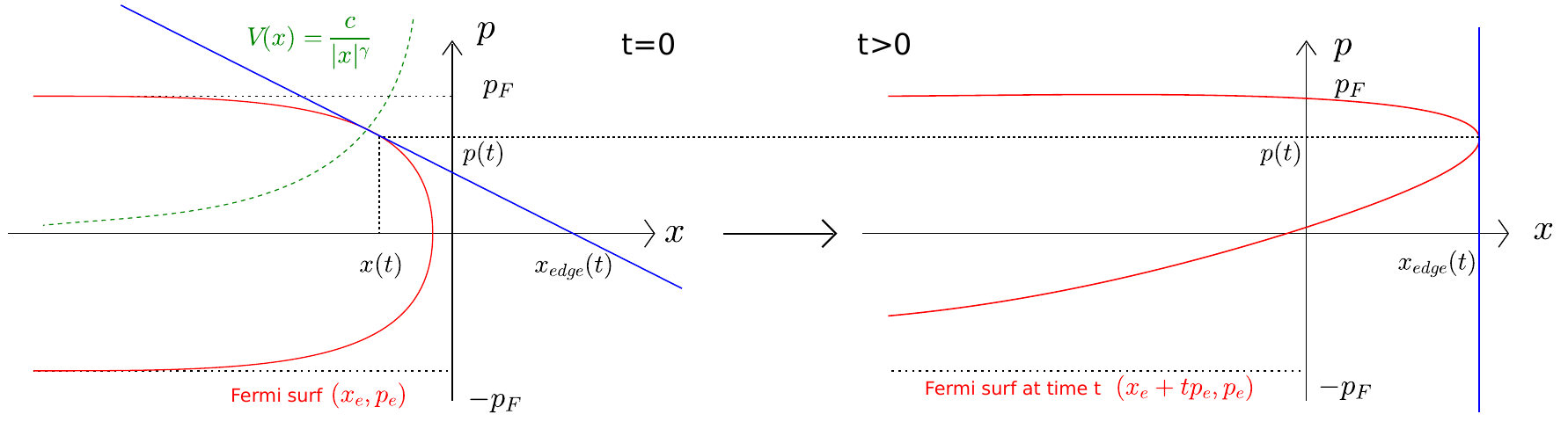}}\captionsetup{width=.85\linewidth}
\caption{Plot of the Fermi surf for the inverse power law potential (red line, left) and its free evolution at time $t$ (red line, right). We also plot the line tangent to the Fermi surf at the quantum front $(x_{edge}(t),p(t))$ (blue, right) and its backward time propagation, i.e., the tangent to the point $(x(t),p(t))$ (blue, left). These two tangent lines appear in \eqref{tangentline}, where in the first line of the equation the integral is performed over a line parallel to the right tangent line, and in the second line of the equation, the integral is performed over a line parallel to the left tangent line. Thus, in the large time limit, the computation of the kernel \eqref{tangentline} probes the edge properties of the Wigner function in the inverse power law potential away from the center of the potential ($x_s\to-\infty$), leading to the Airy kernel at the quantum front \eqref{dynamic_airy}. }
\label{Fig_fermi_surf_dyn}
\end{figure}

{\bf Quench.} Then, we switch-off the trap and consider the free evolution of fermions, producing a positive current, i.e. particles moving from left to right. For free fermions, the dynamical evolution of the Wigner function is simply the Liouville dynamics \al{\partial_t W(x,p,t) + p \partial_x W(x,p,t)=0,} such that
\be\label{liouville_dyn}
W(x,p,t)=W(x-pt,p,t=0)=W_0(x-pt,p),
\ee
where $W_0(x,p)$ is the Wigner function at initial time. This implies that the Fermi surf will be deformed with time (see Fig. \ref{Fig_fermi_surf_dyn}), such that $x(p,t)$ the Fermi surf at time $t$ is given by 
\be
x(p,t)=x_s(p)+pt.
\ee

Therefore, we will observe a quantum front, i.e., the point where the density vanishes in the classical limit, propagating to the right. We look for the position $x_{edge}(t)$ of the propagating quantum front, which is given by $x_{edge}(t) = \max_p {x(p,t)}$. This obeys the equation $\partial_p x(p(t),t) = 0$, defining the corresponding point $p(t)$ and $x(t)=x_s(p(t))$, which is the projection of the quantum front on the Fermi surface of the initial Wigner function. The functions $p(t)$ and $x(t)$ obey the two equivalent equations
\al{\label{Quantum_front}
&x_s'(p(t))+t=0,\\
&p_s'(x(t))+\frac{1}{t}=0.\nn
}
One can verify that $x_{edge}(t)$ is a monotonically increasing function of $t$, such that the front is moving to the right. By construction $x(t)$ and $p(t)$ are monotonous, they have the following large time limit (see Fig. \ref{Fig_fermi_surf_dyn})
\al{&\lim\limits_{t\to\infty}x_{edge}(t)=+\infty,\\
&\lim\limits_{t\to\infty}x(t)=-\infty,\nn\\
&\lim\limits_{t\to\infty}p(t)=p_F,\nn}
and the three of them are linked by the relation
\be\label{xedge}
x_{edge}(t)=x(p(t),t)=x(t)+tp(t).
\ee
Their large time asymptotics are
\al{
& p(t) \underset{t\to\infty}{\simeq} p_F - \frac{1}{(\gamma t)^\frac{\gamma}{\gamma + 1}}(\frac{c}{p_F})^{\frac{1}{\gamma+1}},\\
& x(t) \underset{t\to\infty}{\simeq} -(\frac{\gamma c t}{p_F})^\frac{1}{1+\gamma}\nn
.}
Going back to the kernel, the Liouville dynamics \eqref{liouville_dyn} gives the time dependent kernel at time $t$ as a functional of the initial Wigner function 
\al{\label{tangentline}
& C(x,x',t) = \int dp e^{-i\frac{p}{\hbar}(x-x')}W(\frac{x+x'}{2},p,t)\\
&\quad \quad \quad  = \int dp e^{-i\frac{p}{\hbar}(x-x')}W_0(\frac{x+x'}{2}-pt,p).\nn
}
As illustrated in Fig. \ref{Fig_fermi_surf_dyn}, the integral \eqref{tangentline} giving the correlation at the quantum front $x_{edge}(t)$ is performed over a line parallel to the line tangent to the Fermi surf at the quantum front of the Wigner function at time $t$. In the second line of \eqref{tangentline}, the integral is now performed on a line parallel to the line tangent to the Fermi surf of the initial Wigner function at the point $(x(t),p(t))$. Thus the correlation at the quantum front is dominated by the part of the integral that is close to the Fermi surf. Hence, using the Airy scaling \eqref{Airyscaling1} for the initial Wigner function yields
\al{\label{dumlab}
& C( x_{edge}(t)+y, x_{edge}(t)+y',t)\\
&=\int dp e^{-i\frac{p}{\hbar}(y-y')}W_0((p(t)-p)t+x(t)+\frac{y+y'}{2},p)\nn\\
&=\frac{1}{\hbar t} e^{-i\frac{p(t)}{\hbar}(y-y')}\int\frac{dx}{2\pi}	 e^{i\frac{x-(x(t)+\frac{y+y'}{2})}{\hbar t}(y-y')}\mathcal{W} (a(x)),\nn
}
where the last line is obtained through the change of variable 
\al{x=(p(t)-p)t+x(t)+\frac{y+y'}{2},}
and $a(x)$ is expressed using \eqref{Airyscaling1} as
\al{\label{dumlab2}
& a(x) = \kappa(x)\delta p (x),\\
& \delta p (x) = p(t)-p_s(x)-\frac{x-(x(t)+\frac{y+y'}{2})}{t},\nn\\
&\simeq \frac{(x-x(t))^2}{2}|p_s''(x(t))|+\frac{y+y'}{2t}.\nn
}
The integral in the last line of \eqref{dumlab} is dominated by the vicinity of $x\simeq x(t)$, such that we can use \eqref{Quantum_front} to expand $\delta p(x)$ around $x(t)$ which is done in \eqref{dumlab2}. Now expanding $a(x)$ to the second order in $x$ we have

\al{
&\mathcal{W}(a(x))\simeq\int_0^\infty dz \Ai\left[z+(x-x(t))^2 \left|\frac{p_s''(x(t))}{\hbar}\right|^{2/3}+\frac{y+y'}{t}\left|\hbar p_s''(x(t))\right|^{-1/3}\right.\nn\\
&\left.+(x-x(t))\kappa'(x(t))\frac{y+y'}{2^{1/3}t} + \frac{(x-x(t))^2}{2}\kappa''(x(t))\frac{y+y'}{2^{1/3}t}\right].
}
In order to recover the Airy kernel, we set the new variables 
\al{\label{changeofvar}\begin{cases} \tilde x = (x-x(t))|\frac{p_s''(x)}{\hbar}|^{1/3}\\ \tilde{y} = \frac{y}{t|\hbar^2 p_s''(x)|^{1/3}} \\ \tilde{y}' = \frac{y'}{t|\hbar^2 p_s''(x)|^{1/3}} \end{cases},}
leading to
\al{
&\mathcal{W}(a(x))\simeq\int_0^\infty dz \Ai(z+\tilde{x}^2+ \tilde{y}+ \tilde{y}' + \tilde x (\tilde y+\tilde y')2^{-1/3}\hbar \kappa'(x(t))\\ 
&\quad \quad \quad \quad \quad \quad \quad \quad \quad + \tilde{x}^2 (\tilde y+ \tilde y')\frac{\hbar^{4/3} \kappa''(x(t))}{2^{4/3}(|p_s''(x(t))|)^{1/3}}).\nn
}
The two last terms can be discarded if 
\be\label{expcond55} 
\begin{cases}
 \hbar \kappa ' (x(t)) \ll 1\\
\frac{\hbar^{4/3} \kappa''(x(t))}{(2|p_s''(x(t))|)^{1/3}}   \ll 1
\end{cases} \Leftrightarrow \frac{ \hbar p_F x(t)^{\gamma-1}}{c} \ll 1.
\ee
Notice that this is the same condition as \eqref{expcond5} which is needed for the use of the Airy scaling \eqref{Airyscaling1}. Hence we need to make sure that this condition is verified if we want to observe the Airy kernel. Finally, we end up with the simple form
\be
\mathcal{W}(a(x))\simeq 2 \int_0^\infty dz \Ai(\tilde{x}^2+ \tilde{y}+ \tilde{y}'+2z),
\ee
such that taking into account the change of variable \eqref{changeofvar}, the correlation at the quantum front is given by

\al{
& C( x_{edge}(t)+y, x_{edge}(t)+y',t)=-\frac{1}{t}\frac{1}{(\hbar^2|p_s''(x(t))|)^{1/3}}e^{-i\frac{p(t)}{\hbar}( y-  y')}\\
&e^{-\frac{i}{2\hbar}(y^2-y'^2)}\int_0^\infty dz \int\frac{d\tilde x}{\pi}	 e^{i\tilde x (\tilde y +z- (\tilde y ' +z))}  \Ai(\tilde{x}^2+ \tilde{y} + z + \tilde{y}' + z).\nn
}
We end the computation using the integral identity \eqref{book_formula}
which leads to 


\al{\label{dynamic_airy}
&C(x,x',t)\simeq -\frac{1}{w(t)} e^{-\frac{i}{\hbar}( p(t)(x-x')+\frac{(x-x_{edge}(t))^2}{2t}-\frac{(x'-x_{edge}(t))^2}{2t})}\\
& \quad \quad \quad \quad \quad \quad \times K_{\Ai}(\frac{x-x_{edge}(t)}{w(t)},\frac{x'-x_{edge}(t)}{w(t)}),\nn
}
that is the emergence of the Airy kernel at the quantum front as stated in \eqref{result2}.

\section{Discussion}\label{Conclusion}

In this article, we were interested in the ground state Wigner function of a noninteracting Fermi gas with the inverse power law potential. We analyzed the fine structure of the edge of the Wigner function called the Fermi surf in the semiclassical limit, aiming to determine whether it exhibits the smooth Airy scaling \eqref{Airyscaling}. We differentiated between two regimes. The first one, "near the gas edge", exhibits the Airy scaling unless the condition \eqref{nearcond} is violated. This occurs when the indirect classical path induced by the repulsive inverse power law cannot be neglected. The second regime, "away from the gas edge", is valid unless the condition \eqref{expcond5} is violated.\par

In the second part of the article, we proposed a quench setup where the system is initialized in the previously studied ground state of the inverse power law potential. Then, the potential is turned off, and the fermions evolve freely. Using the first part, we were able to prove the emergence of the Airy kernel at the edge of the propagating front. Additionally the gas exhibits anomalous diffusion around the quantum front with generalized diffusion exponent ranging from $\frac{1}{3}$ to $\frac{2}{3}$ depending on the exponent $\gamma$ of the pre-quench potential.\par 

This results raise new axis of research. The Airy kernel arises as a consequence of the initial equilibrium Wigner function, hence it would be interesting to see if changing the initial state by some thermal state would lead to the kernel observed at the edge of a thermal Fermi gas. Additionally, it would be interesting to observe this phenomenon in an experiment, even though this seems to be challenging given the current spatial resolution of cold atoms experiments. Such a quench could then be used as a method to probe fine details of the Fermi surf of the pre-quench Wigner function.
\vspace*{.5cm}

\noindent{\bf Acknowledgements} We thank P. Le Doussal and G. Schehr for helpful discussions. We thank B. Douçot and P. Vignolo for useful comments. We acknowledge support from ANR grant ANR-17-CE30-0027-01 RaMaTraF.

\begin{appendix}

\section{The inverse squared potential \texorpdfstring{$\gamma=2$}{Lg}}\label{app:gamma2}
Let us give details on the behaviour of the Wigner function of the gournd state of the inverse square potential. The single particle problem of the inverse square potential can be solved exactly, and the single particle eigenfunctions are simply related to the Bessel function $J_{\nu}$ through.
\be
\psi_p(x)=\sqrt{\frac{px}{\hbar}}J_{\nu}(\frac{px}{\hbar}), \quad J_\nu(x) = \frac{1}{2\pi} \int_{-\pi}^\pi d\tau e^{i(\nu \tau - x \sin(\tau))},
\ee
with $\nu=\sqrt{\frac{1}{4}+\frac{2c}{\hbar^2}}\simh \frac{\sqrt{2c}}{\hbar}$. For now we will keep $\nu=\frac{\sqrt{2c}}{\hbar}$. This leads to the following expression for the Wigner function

\al{
&W(x,p)=\frac{1}{2\pi \hbar^3}\int_0^{p_F} p'dp' \int_{-2x}^{2x} dy e^{\frac{i \nu y}{x_\alpha(p)}} \sqrt{(x+\frac{y}{2})(x-\frac{y}{2})}	 \\
&\quad \quad \quad \quad \times J_\nu(\nu\frac{x+y/2}{x_\alpha(p')}) J_\nu(\nu\frac{x-y/2}{x_\alpha(p')}).\nn
} 
Here $x_\alpha(p')=\frac{\sqrt{2c}}{p'}$ is the turning point for a wave of impulsion $p'$, i.e. the point where $\frac{p'^2}{2}=V(x_\alpha)$. This leads to

\al{\label{saddlepointform1}
&W(x,p)=\frac{1}{(2\pi \hbar)^3}\int_0^{p_F} p'dp' \int_{-2x}^{2x} dy  \int_{-\pi}^\pi d\tau_1 d\tau_2\sqrt{x^2-\frac{y^2}{4}}  e^{i\nu F (\tau_1,\tau_2,p',y)},\nn\\
&F(\tau_1,\tau_2,p',y)=
\tau_1+\tau_2 - p'\frac{x+\frac{y}{2}}{\sqrt{2c}} \sin(\tau_1)- p'\frac{x-\frac{y}{2}}{\sqrt{2c}} \sin(\tau_2)+\frac{py}{\sqrt{2c}}.
}
This is a four dimensional integral with a  saddle point. Let us analyse the integral limit at large $\nu$ with the saddle point method. The saddle point equation $\nabla F=0$ for $x>0,p>0$ leads to the unique critical point $(\tau_1^*,\tau_2^*,p^*,y^*)$
\al{
&\tau_1^*=-\tau_2^*=\arcsin(p/p^*),\nn\\
&\frac{(p^*)^2}{2}=\frac{p^2}{2}+\frac{c}{x^2},\\
&y^*=0.\nn
}
We voluntary forgot the solution $\tau_1=-\pi -\arcsin(\sin(\tau_2)+\frac{2p}{p'})$, and $\tau_1=\pi-\arcsin(\sin(\tau_2)+\frac{2p}{p'}) $ which arise if the conditions $x>0,p>0$ are relaxed. Additionally we have $F(\tau_1^*,\tau_2^*,p^*,y^*)=0$ leading to the second order expansion
\al{\label{relation1}
F(\tau_1,\tau_2,p',y) \simeq \frac{xp}{\sqrt{2c}}\frac{\tau_1'^2-  \tau_2'^2}{2}+\frac{(\tau_1' +\tau_2')  p''}{p^*}+ y(\frac{p}{\sqrt{2c}p^*} p'' +\frac{\tau_2'- \tau_1'}{2x}),
}
with $ \tau_{i}' =\tau_i-\tau_i^*,\quad  p''=-(p'-p^*)$.  Notice that the $y$ variable plays the role of a Dirac delta function with the change of variable $y=\hbar \tilde y$. This could already be seen in \eqref{saddlepointform1} by isolating the integral over $y$ which can be expressed as a Bessel function and converges towards a Dirac delta function when $\hbar$ is small. This delta Dirac impose the following relation 
\be\label{deltacondition0}
p+\frac{p'}{2}(\sin(\tau_2)-\sin(\tau_1))=0,
\ee
which after a first order expansion gives
\be\label{deltacondition1}
\frac{p}{\sqrt{2c}p^*} p'' +\frac{ \tau_2' - \tau_1'}{2x}=0.
\ee
Substituting this in Eq. \eqref{relation1} results in

\be
F(\tau_2,p') \simeq \frac{\sqrt{2(p^2/2+V(x))}}{V(x)}  \tilde \tau_2 \tilde p + \frac{\sqrt{2}p}{V^{3/2}(x)}\frac{\tilde p ^2}{2}.
\ee
We see that the second order term $\tau_1'^2$ and $\tau_2'^2$ from Eq. \eqref{relation1} are cancelled. Hence we need to expand $F$ to third order 
\al{\label{relation2}
&F(\tau_2,p') \simeq \frac{xp}{\sqrt{2c}}(1-\frac{ p''}{p^*})\frac{ \tau_1'^2- \tau_2'^2}{2}+\frac{( \tau_1' + \tau_2' )  p''}{p^*}+\frac{ \tau_1'^3 + \tau_2'^3}{6}\\
&+ y(\frac{p}{\sqrt{2c}p^*} p'' +\frac{ \tau_2 '-  \tau_1'}{2x}+ \frac{ p''}{2 x p^*}( \tau_1' - \tau_2' ) +\frac{p}{4\sqrt{2c}}( \tau_1'^2 +  \tau_2'^2)).\nn
}
Once again, we need to enforce the constraints from the delta Dirac, reducing $F$ to a two dimensional function. This is done by expanding \eqref{deltacondition0} to the second order such that $\frac{p}{\sqrt{2c}p^*} p'' +\frac{ \tau_2 '-  \tau_1'}{2x}+ \frac{ p''}{2 x p^*}( \tau_1' - \tau_2' ) +\frac{p}{4\sqrt{2c}}( \tau_1'^2 +  \tau_2'^2)=0$. This can be inverted as
\al{
&  \tau_1' =  \tau_2 ' + \frac{2 xp}{\sqrt{2c}}\frac{ p''}{p^*} \\
& + \frac{px}{\sqrt{2c}}  \tau_2'^2 + \frac{x^2p^2}{c} \frac{ p''  \tau_2'}{p^*} + (\frac{2 xp}{\sqrt{2c}}+2 (\frac{xp}{\sqrt{2c}})^3)\frac{ p''^2}{(p^*)^2}.\nn
}
Finally the constraint is substituted back in \eqref{relation2} leading to 
\al{\label{Fthirdorderexpand}
&W(x,p)=\frac{1}{(2\pi \hbar)^{2}}\frac{2x^2 p^*}{\sqrt{2c}} \int_{p^*-p_F}^{p^*} dp''\int_I d\tau_2e^{i\nu F(\tau_2,p')},\nn\\
&F(\tau_2,p') \simeq (1+3\frac{p^2/2}{V(x)}) \frac{\tau_2'^3}{3} + \frac{\sqrt{2(p^2/2+V(x))}}{V(x)}  \tau_2' p'',\\
& + \frac{\sqrt{2}p}{V^{3/2}(x)}\frac{ p''^2}{2}.\nn
}
Note that the delta Dirac function brings an additional factor $2\pi \hbar \frac{2x^2}{\sqrt{2c}}$ in front of the integral, and $I=\{\tau'\in [-\pi,\pi]|\exists \tau,\quad p+p'\frac{\sin(\tau')-\sin(\tau)}{2}=0\}$.
Proceeding to the change of variable
\al{
\begin{cases}
\tau_2'= \nu^{-1/3}(1+3\frac{p^2/2}{V(x)})^{-1/3}\tilde{\tau}\\
p''= \nu^{-2/3}(1+3\frac{p^2/2}{V(x)})^{1/3} \frac{V(x)} {\sqrt{2(p^2/2+V(x))}} \tilde{p}=\frac{\hbar^{2/3}}{2\sqrt{2}}\frac{(p^2V''(x)+(V'(x))^2)^{1/3}}{\sqrt{\frac{p^2}{2}+V(x)}}  \tilde{p}
\end{cases},
}
and for fixed $x$, using the relation 
\al{(p^*-p_F)\frac{2p^*}{\hbar^{2/3}(p^2V''(x)+(V'(x))^2)^{1/3}}=2^{2/3} a \frac{2 p^*}{p^*+p_F},}
leads to the Airy scaling \eqref{Airyscaling} 

 \hspace{1cm}
 
\noindent{\bf Large $x$} In section \ref{art42}, we study the emergence of the Airy kernel at the propagating front \eqref{result2}. This depends on the behavior of the initial Wigner function at position $x(t)$, i.e. for $|x|\to+ \infty$. Therefore, we need to now if the Airy scaling is still verified when $|x|\to+\infty$. For fixed $x$, the Airy scaling is verified only because as $\hbar$ goes to 0, the last term of \eqref{Fthirdorderexpand} also goes to 0, and the interval $I$ goes to $\mathbb{R}$. However for $x$ large, the last term of \eqref{Fthirdorderexpand} (and also all other third order term of expansion of $F$) is a power of $\frac{x p \hbar}{c}$. In fact even the interval $I$ has a width of order $\Delta \simeq (\frac{x p \hbar}{c})^{-1/3}$.
Therefore, for the inverse squared potential, the Airy scaling is recovered only under the condition
\be\label{expcond11}
\frac{xp\hbar}{c}\ll 1.
\ee
This implies that $\hbar$ has to be small and $x$ cannot be taken too large.

\section{Short time propagator expansion}\label{app:inversepowerlaw}
Here we give some details on the short time propagator expansion presented in the appendix of \cite{dean2018wigner}. We carefully restore the $\hbar$ factors in order to perform the small $\hbar$ analysis.\par
 \hspace{1cm}
 
{\bf General smooth potential.} 
We consider the Wigner function written as \eqref{toexpand} at a point near the Fermi surf $(x,p)\simeq (x_s,p_s)$. The small time expansion of Eq. \eqref{toexpand} is valid if the arguments of $V$ in Eq. \eqref{wigner_prop} are small compared to $x\simeq x_s$. In other words, we require that the following conditions hold:
\al{\label{expcond2}
\begin{cases}
\frac{p_s t}{m x_s} \ll 1\\
\frac{\hbar t}{m x_s^2} \ll 1
\end{cases}.}
Considering the Brownian bridge, the expansion to order 3 results in
\al{\label{order3toexp}
& S(x,p) \simeq - \frac{t^2}{m} (\frac{1}{24}+\frac{1}{12}) V''(x_s)+ \frac{t^3}{24 m \hbar }(V'(x_s))^2 \\
&   -\frac{t^3}{m^2}[-\frac{1}{24 \hbar}p_s^2 V''(x_s) + \hbar (\frac{1}{640}+\frac{1}{480}+\frac{1}{240})V''''(x_s)]+ O(t^4).\nn
}
because we want to recover the Airy function, we want the second order to be negligible. Let us assume that, finally, the only remaining term is
\be
S(x,p)\simeq \frac{t^3}{24 m \hbar }(\frac{p_s^2}{m} V''(x_s)+V'(x_s)^2).
\ee
Then we can make the change of variable $t=2^{2/3} \hbar \tau/e_N $ in the integral \eqref{toexpand}, with 
\al{e_N=\hbar /t_N=\frac{\hbar^{2/3}}{(2m)^{1/3}}(\frac{p_s^2}{m} V''(x_s)+V'(x_s)^2)^{1/3},}
 and the Airy scaling emerges (see Eq. \eqref{airy_scale_concle}). If we assume that the integral is dominated by $t \sim t_N$, we obtain the condition for neglecting the discarded terms in \eqref{order3toexp} as
\al{\label{expcond3}
 &\frac{t^2}{m} V''(x_s)\simeq \frac{t_N^2V(x_s)}{m x_s^2} \ll 1 ,\\
 &\hbar V''''(x_s) \ll \frac{1}{\hbar}	p_s^2 V''(x_s) + \frac{m}{\hbar} V'(x_s)^2\simeq t_N^{-3}.\nn
}
 Hence, the Airy sclaing is valid if the conditions \eqref{expcond2} and \eqref{expcond3} are satisfied.
 \hspace{1cm}
 
{\bf Conditions for the inverse power law potential.} Now we apply the short time expansion to the special case of the inverse power law potential. This is done by checking in which regime \eqref{expcond2} and \eqref{expcond3} are satisfied.\par
 For the part of the Fermi surf near the edge of the gas (see Fig. \ref{fig:wigner_regime}) we parametrize the Fermi surf as 
 \al{&x_s=\bar x x_{edge},\\
 &p_s=\sqrt{2\mu(1-\bar x^{-\gamma})},\nn}
 where we recall that $x_{edge}=(\frac{c}{\mu})^{1/\gamma}$.
 With this scaling, the conditions \eqref{expcond2}, and \eqref{expcond3} lead to
 \al{\label{omega1}&\frac{p_s t_N}{x_s}\simeq (\sqrt{\omega}\bar{x}^{\gamma-1})^{1/3}\ll 1,\\
 &\frac{\hbar t_N}{x_s^2}\simeq (\omega^2 \bar{x}^{\gamma-4})^{1/3} \ll 1,\nn\\
 &t_N^2 V''(x_s)\simeq \left( \frac{\omega}{\bar{x}^{\gamma+2}}\right)^{1/3}\ll 1,\nn\\
 &\hbar V''''(x_s)t_N^3 \simeq \frac{\omega}{\bar{x}^2}\ll1,\nn}
 where $\omega=\frac{\hbar^2}{\mu^{\frac{\gamma-2}{\gamma}}c^{\frac{2}{\gamma}}}$ such that near the gas edge (that is $\bar{x}$ of order one), the expansion of Eq. \eqref{toexpand} and hence the Airy scaling is valid if the single condition 
 \al{ \omega=\frac{\hbar^2}{\mu^{\frac{2-\gamma}{\gamma}}c^{\frac{2}{\gamma}}} \ll 1, }
 is satisfied, which yields \eqref{nearcond}. The main commentary to do here is that we need either $\hbar$ to be small or, if $\gamma>2$, we need $\mu$ to be large.\par
 
Now, we examine the part of the Fermi surf, away from the edge of the gas (see Fig. \ref{fig:wigner_regime}), that is for $p\sim \sqrt{m\mu}$ and $x_s$ as large as we want. First taking $\bar{x}\to \infty$ in \eqref{omega1} we see that the third and fourth line of \eqref{omega1} which corresponds to \eqref{expcond3} are always satisfied. Therefore, we can focus on the two first lines of \eqref{omega1}, that is on \eqref{expcond2}. This yields
  \al{\label{expcond4}
 & \frac{p_s t_N}{ x_s} \simeq \left(\frac{ \hbar \mu^{1/2} x_s^{\gamma-1}}{c}\right)^{1/3} \ll 1, \\
  & \frac{\hbar t_N}{ x_s^2} \simeq \left(\frac{\hbar^{4} x_s^{\gamma-4}}{ c \mu}\right)^{1/3}\ll 1 .\nn
}
 Because the first condition is more restrictive than the second one, in the limits of interest ($\hbar \to 0$ or $\mu \to \infty$ with $x_s$ large), the Airy scaling is correct only if the second condition is satisfied
\be
\frac{\hbar \sqrt{\mu} x_s^{\gamma-1}}{c}\ll 1,
\ee
which is \eqref{expcond5}. The conclusion is that $\mu\to\infty$ cannot satisfy the required condition for the emergence of the Airy scaling \eqref{Airyscaling}. However, this is possible at small $\hbar$, if $\gamma<1$ on the complete Fermi surf, and if $\gamma>1$ only for $x_s<\hbar^{1/(1-\gamma)}$.


\end{appendix}

\printbibliography[heading=bibintoc]

\end{document}